\documentclass[letterpaper, 10pt, conference]{ieeeconf} 
\usepackage{amsmath,graphicx}
\usepackage{caption}
\usepackage{array}
\usepackage{booktabs}

\newcolumntype{C}[1]{>{\centering\let\newline\\\arraybackslash\hspace{0pt}}m{#1}}
\graphicspath{{images/}}
\IEEEoverridecommandlockouts
\title{\LARGE \bf Ultrasound segmentation using U-Net: learning from simulated data and testing on real data}
\author{Bahareh Behboodi$^{1}$ and Hassan Rivaz$^{2}$
\thanks{*This work was supported by the Richard and Edith Strauss Foundation.}
\thanks{$^{1}$Bahareh Behboodi is with the Department of Electrical and Computer Engineering, Concordia University, Canada
        {\tt\small b\_behboo@encs.concordia.ca}}%
\thanks{$^{2}$Hassan Rivaz is with the Department of Electrical and Computer Engineering and PERFORM Center, Concordia University, Canada
        {\tt\small hrivaz@ece.concordia.ca}}%
}

\begin{document}

\maketitle
\begin{abstract}
Segmentation of ultrasound images is an essential task in both diagnosis and image-guided interventions given the ease-of-use and low cost of this imaging modality. As manual segmentation is tedious and time consuming, a growing body of research has focused on the development of automatic segmentation algorithms. Deep learning algorithms have shown remarkable achievements in this regard; however, they need large training datasets. Unfortunately, preparing large labeled datasets in ultrasound images is prohibitively difficult. Therefore, in this study, we propose the use of simulated ultrasound (US) images for training the U-Net deep learning segmentation architecture and test on tissue-mimicking phantom data collected by an ultrasound machine. We demonstrate that the trained architecture on the simulated data is transferrable to real data, and therefore, simulated data can be considered as an alternative training dataset when real datasets are not available. The second contribution of this paper is that we train our U-Net network on envelope and B-mode images of the simulated dataset, and test the trained network on real envelope and B-mode images of phantom, respectively. We show that test results are superior for the envelope data compared to B-mode image.

\end{abstract}

\section{Introduction}
\label{sec:intro}
Ultrasound (US) is a non-invasive diagnosis methodology since it utilizes low energy US waves in order to capture tissue characterizations. US image segmentation has been used in diagnosis as well as image-guided interventions \cite{ref4}. Despite the wide applications of US images, they are largely contaminated with speckle noise, which renders accurate segmentation of different structures challenging. Thus, recently, US segmentation problems have attracted a growing number of research groups, which has led to important achievements using both traditional machine learning techniques and more recently deep learning algorithms \cite{ref4, ref3}. 

Recent deep learning techniques have shown promising results on various data analysis tasks due to their power in learning efficient and compact data representations in classification \cite{ref1}, segmentation \cite{ref2}, and automatic speech recognition \cite{ref10} tasks. Despite this noticeable success, training deep learning architectures requires large amount of data. In case of US imaging, data acquisition and interpretation by clinical experts are very expensive. To address this issue, two research avenues are often pursued. First, an architecture can be used that is designed with this limitation in mind. The U-Net architecture \cite{ref3}, built upon fully convolutional network \cite{ref2}, is the most well-known architecture for biomedical image segmentation which takes advantages of several convolutional, max-pooling, and upsampling layers. The application of U-net on simulated US images has been recently proposed \cite{ref7}. Second, in order to increase the size of data, various data augmentation strategies for medical images have been proposed \cite{ref8, ref9}. However, to the best of our knowledge, there are no applications of training a deep neural network architecture on the simulated US data and test on the real data. In addition, there is no previous work that compares the utility of envelope data and B-mode images.
Therefore, we address the following two questions in this work:
\begin{itemize}
\item Can a network be trained on simulation data and tested on real data?
\item Which data provides the best segmentation results, envelope data or B-mode images?
\end{itemize}

Radio-frequency (RF) data is the raw data that is generated from beamforming the channel data. This data is not suitable for visualization as it contains very high frequencies, and therefore its envelope is extracted. Since the envelope data has very high dynamic ranges, it undergoes a logarithmic compression, which is often called B-mode US image and is suitable for display. Although logarithmic compression lead to better visualization of the US data, pixels with higher values are compressed and a lot of information is lost during this compression transform. In this work, we compare the segmentation results of envelope data and B-mode image.	

Herein, we propose a novel strategy based on the use of simulated US images as the training set. We utilize the U-Net architecture to learn the segmentation masks of the simulated US images. Afterward, the segmentation masks of tissue-mimicking phantom data is predicted using the trained network. The outline of this study is as follows. In Section \ref{sec:method}, details of U-Net architecture along with its parameters, as well as simulated and phantom data used for training and testing, respectively, are elaborated. Section \ref{sec:results} presents the predicted segmented masks achieved by implementing our proposed strategy as well as quantitative comparison of the results. We conclude this paper in Section \ref{sec:conclusion} with some directions for future research.

\section{Method}
\label{sec:method}

\subsection{U-Net Architecture}
\begin{figure}[h!]
  \centering
  \centerline{\includegraphics[width=8.5cm]{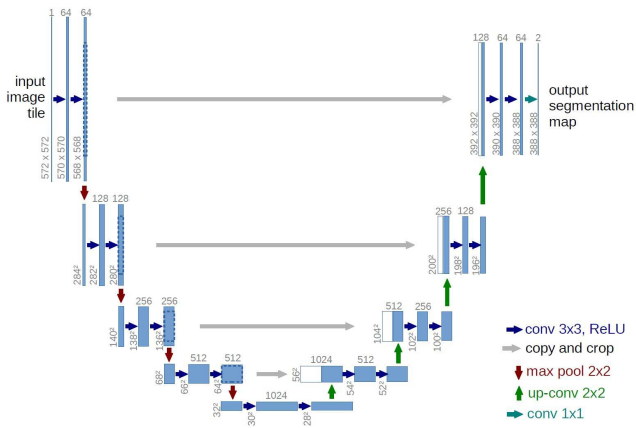}}
  \caption{U-Net architecture \cite{ref3}}
  \label{fig:unet}
\end{figure}

U-Net architecture consists of two paths, a contracting path and an expansive path. The contracting path (left path in Fig. \ref{fig:unet}) consists of several repetition of two convolution (conv) and one max-pooling (max pool) layers with the kernel sizes of $3\times3$ and $2\times2$, respectively. The expansive path (right path in Fig. \ref{fig:unet}) comprises of the repetition of concatenation of the features extracted from corresponding layers in contracting path (copy and crop), two convolution, and one upsampling (up-conv) layers. In this path, the kernel for convolution layer is same as the contracting path and $2\times2$ for upsampling layer. The final layer has one $1\times1$ convolution kernel. 

Activation functions in convolution layers are set to rectified linear units (ReLU) \cite{ref11} except the last layer, which is set to Softmax. We pose the segmentation problem as a pixel-wise classification that lead to three class classification for our dataset. In the last layer of our architecture three $1\times1$ convolution kernels are used, and the loss function is set to categorical cross-entropy. The learning rate, optimizer, and weight initializer function are set to 0.00001, Adam \cite{ref12}, and He-normal \cite{ref13}, respectively. For the remaining parameters, we follow the initial parameters proposed in \cite{ref3}. U-Net is trained through 100 epochs with batch size of 8 in order to fit data in the memory. The codes for implementing U-Net are scripted on python (version 3.6) using Keras with Tensorflow backend. A Titan Xp NVIDIA GPU with 12 GB of memory on Ubuntu 16.04 LTS is used for training and testing. 

\subsection{Simulation Data}
\label{subsec:sim}
Simulated RF is generated using the publicly available ultrasound simulation software, Field\_II \cite{ref5,ref6} based on MATLAB release 2018, followed by envelope extraction. The virtual ultrasound transducer parameters are initialized as outlined in Table \ref{tab:virtrans}. We randomly distribute scatterers in all phantoms to ensure each $mm^3$ has in average 4 scatterers. The simulated images randomly consist of hyperechoic lesions (i.e. tissues with higher echogenicity), and anechoic lesions (i.e. tissues with lower echogenicity). In hyperechoic lesions, the scatterers intensities are $k$ times larger than background in which $k$ is a random integer value between 1 and 10. The lesions are placed between -20 to +20 $mm$ in the lateral direction and between 30 to 90 $mm$ in the axial direction. Lesion shapes are circles or ellipses with random sizes. The radii of circles are between 1-3 $mm$, and semi-major and semi-minor axes of ellipses are between 5-9 and 1-5 $mm$, respectively. In total, 700 images are simulated. We then split the data to training, validation and testing data sets considering 60\%, 15\%, and 25\% splitting factors of the total images, yielding 420, 105, and 175 images, respectively. Envelope and B-mode images with initial size of $14,069\times50$ are then resized to $512\times512$, mirrored (with the same mirroring factor as described in \cite{ref3}) to size $572\times572$. The intensity range is normalized to the range between 0-1 before feeding to the U-Net. As mentioned earlier, a simulated data may consist of one hyperechoic and one anechoic lesion. Therefore, including the background, three classes should be categorized. As such, the ground truth of a simulated image is in the size of $388\times388\times3$.
\begin{table}[h]
	\centering
	\caption{Parameters of virtual ultrasound transducer.}
	\label{tab:virtrans}
	\small
	\begin{tabular}{ m{3.7cm}  C{3.7cm}}
		\toprule[0.5mm]
		\multicolumn{1}{c}{\bf Property Name} & \multicolumn{1}{c}{\bf Property Value} 				\\
		\toprule[0.5mm]
		Number of RF lines		& \multicolumn{1}{c}{ 50}\\
		\midrule
		Start depth of simulation data		& 30 $mm$ from the transducer surface\\
		\midrule
		Depth of simulation data		& 90 $mm$ from the transducer surface\\
		\midrule
		Lateral distance of simulation data	& 40 $mm$ (from -20 to 20 $mm$)\\ 
		\midrule
		Speed of sound		& 1540 m/s\\ 
		\midrule
		Center frequency	& 3.5 MHz\\ 
		\midrule
		Sampling frequency	& 100 MHz\\ 
		\midrule
		\midrule
	\end{tabular}
\end{table}

\subsection{Tissue-Mimicking Phantom Data}
\begin{figure}[h!]
  \centering
  \centerline{\includegraphics[width=6cm]{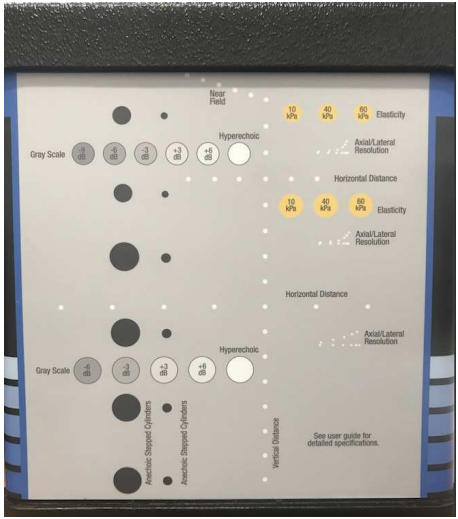}}
  \caption{CIRS Multi-Purpose Multi-Tissue ultrasound phantom}
  \label{fig:phantom}
\end{figure}

US data is acquired from a CIRS Multi-Purpose Multi-Tissue ultrasound phantom with an Alpinion E-Cube system (Bothell, WA) using the L3-12H transducer at the center frequency of 10 MHz and a sampling rate of 40 MHz. Table \ref{tab:phtrans} and Fig. \ref{fig:phantom} indicate the transducer and phantom details, respectively. The phantom data includes different types of lesions in different depths with circular shapes. In this work, total of 6 phantom images with the depth of 40 $mm$ are acquired from different locations of the phantom.
\begin{table}[h]
	\centering
	\caption{Parameters of Alpinion machine in real experiments.}
	\label{tab:phtrans}
	\small
	\begin{center}
	\begin{tabular}{ m{3.7cm}  C{3.7cm}}
		\toprule[0.5mm]
		\multicolumn{1}{c}{\bf Property Name} & \multicolumn{1}{c}{\bf Property Value} 				\\
		\toprule[0.5mm]
		Number of RF lines		& 384\\
		\midrule
		Start depth of simulation data		& 4 $mm$ from the transducer surface\\
		\midrule
		Depth of simulation data		& 40 $mm$ from the transducer surface\\
		\midrule
		Lateral distance of simulation data	& 40 $mm$ (from -20 to 20 $mm$)\\ 
		\midrule
		Speed of sound		& 1540 m/s\\ 
		\midrule
		Center frequency	& 10 MHz\\ 
		\midrule
		Sampling frequency	& 40 MHz \\ 
		\midrule
		\midrule
	\end{tabular}
	\end{center}
\end{table}

\subsection{Quantitative Evaluation}
To evaluate the performance of the network, here we report two different metrics in order to compare the predicted mask with the ground truth mask, Dice Similarity Coefficient (DSC) and $F\textsubscript{2}$ score defined as below:
\begin{equation}
DSC = \frac{2 |P\cap R|}{|P|+|R|} = \frac{2 TP}{2 TP + FP + FN}
\end{equation}
\begin{equation}
 F\textsubscript{2} = \frac{5 TP}{5 TP + 4 FN + FP}
\end{equation}
where $TP$, $FP$, and $FN$ are true positive, false positive, and false negative, respectively. For the simulation data, the images are simulated based on the predefined information of the location of the lesions, which is considered as the ground truth. However, for the phantom data, the ground truth is manually obtained using the ImageJ software \cite{ref14}. 

U-Net is trained on simulated envelope and B-mode images of solely the simulation experiments, yielding two different trained weights. Subsequently, the trained weights are used to test on simulated (different from the training simulation data) and real phantom data, yielding predicted segmentation masks.
 
\section{Results}
\label{sec:results}
To provide a comprehensive comparison, the results of predicted masks derived from training U-Net based on simulated envelope and B-mode images and testing on both simulated and phantom data is illustrated in this section. Furthermore, we outline the $DSC$ and $F\textsubscript{2}$ scores for simulated and phantom data.

\subsection{Predicted Masks for Simulation Data}
\begin{figure}[h!]
 \centering
\begin{minipage}{1\linewidth}
 \centering
 \centerline{\includegraphics[width=9cm]{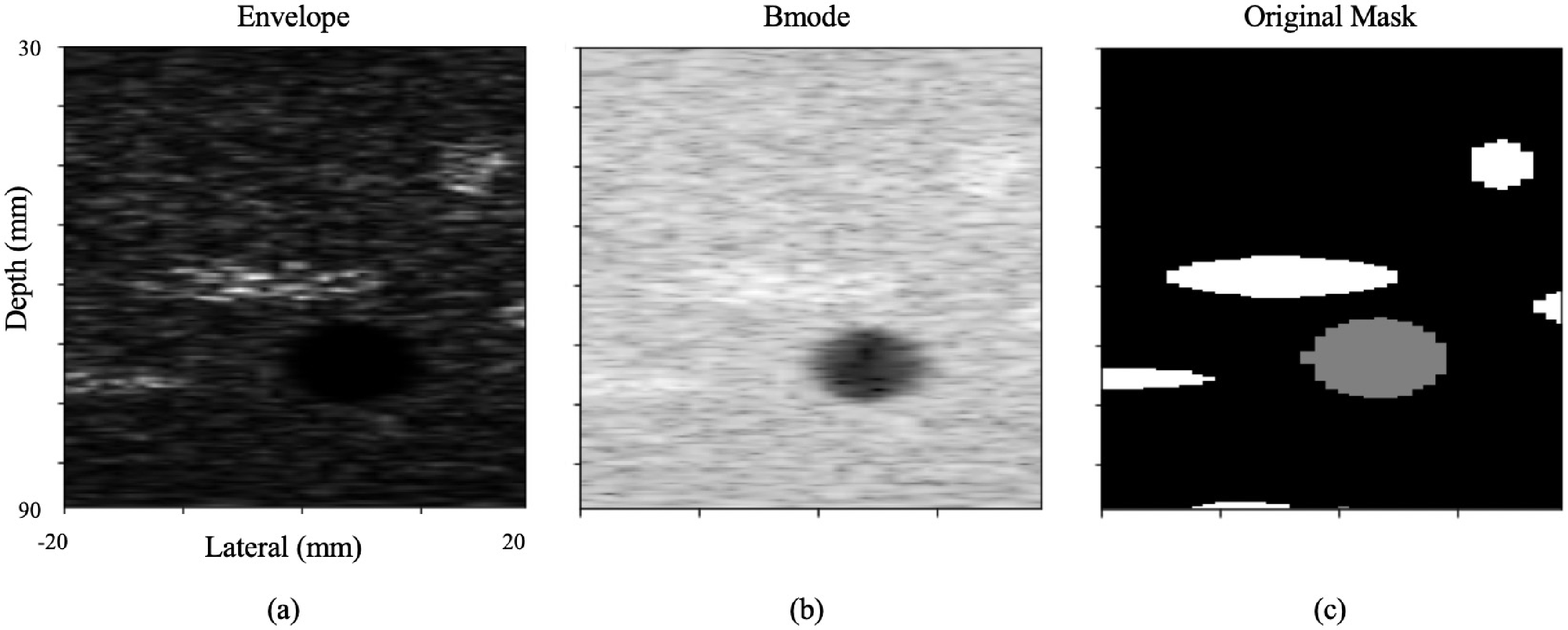}}
\end{minipage}

\begin{minipage}{1\linewidth}
 \centering
 \centerline{\includegraphics[width=6.3cm]{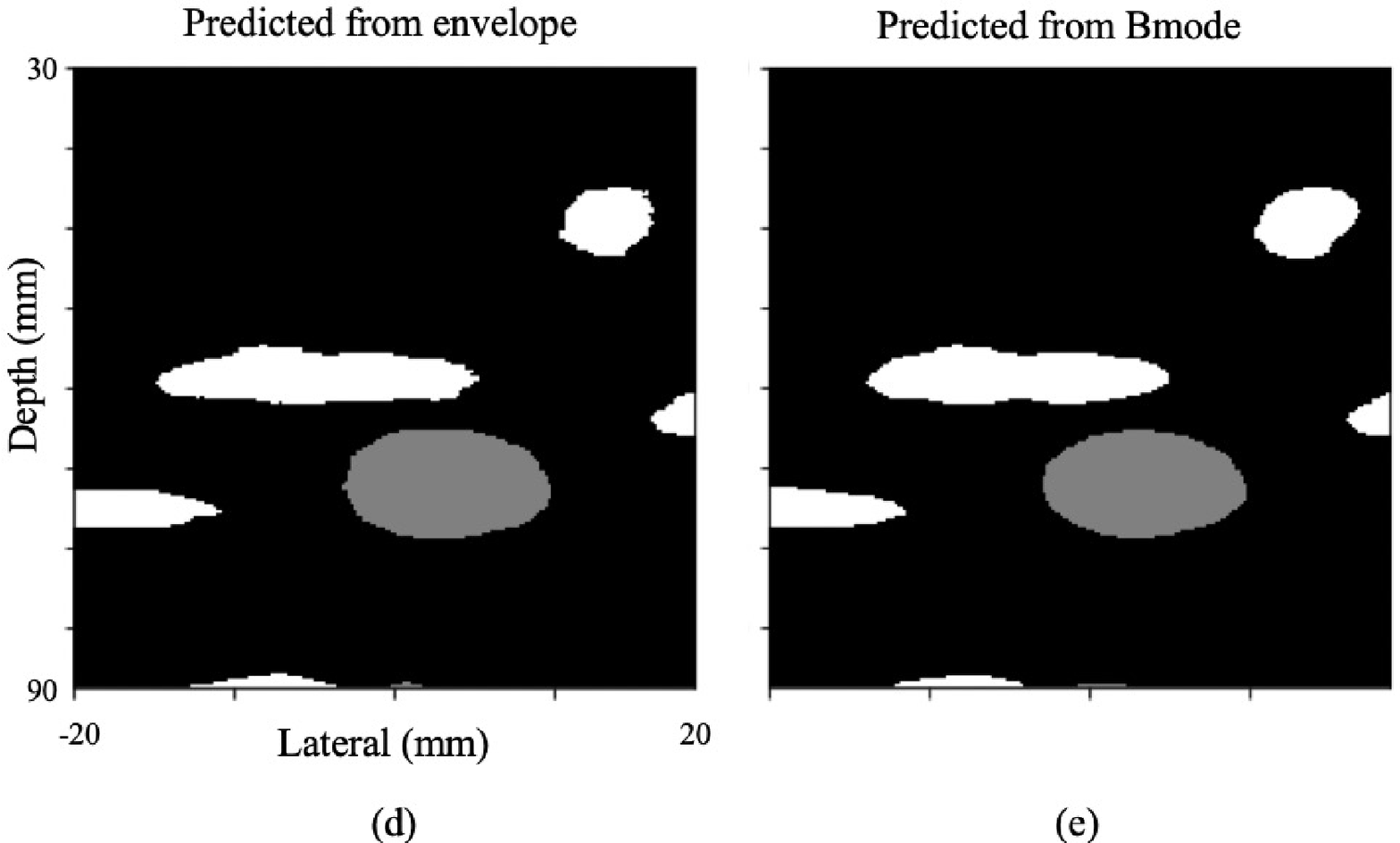}}
\end{minipage}
 \caption{Field\_II simulation images. The first row is an example of (a) envelope data, (b) B-mode image and (c) the ground truth mask. Predicted masks using (d) envelope and (e) B-mode image are shown in the second row.}
 \label{fig:sim}
\end{figure}

Fig. \ref{fig:sim} represents an example of simulated envelope data, B-mode image, the ground truth mask, and the predicted masks. In this particular example, the simulated data consists of 6 types of lesion including 5 hyperechoic and one anechoic. It is important to highlight that four of the lesions are located on the borders and therefore are only partly contained in the image. The predicted masks provide clearer boundaries of all aforementioned lesions compared to the ground truth mask.

Mean and standard deviation of $DSC$ and $F\textsubscript{2}$ scores for predicted masks from the network trained on envelope and B-mode image are summarized in Table \ref{tab:simscore}.

\begin{table}[h]
	\centering
	\caption{$DSC$ and $F\textsubscript{2}$ scores for simulated data.}
	\label{tab:simscore}
	\small
	\begin{tabular}{ m{4cm}  C{1.5cm} C{1.5cm}}
		\toprule[0.5mm]
		\multicolumn{1}{c}{\bf Predicted Mask} & \multicolumn{1}{c}{$\bf DSC$} & \multicolumn{1}{c}{ $\bf F\textsubscript{\bf2}$}\\
		\toprule[0.5mm]
            	Envelope data & $0.85 \pm 0.16$ & $0.87 \pm 0.19$\\
		\midrule
		B-mode image & $0.85 \pm 0.16$ & $0.85 \pm 0.2$\\
		\midrule
		\midrule
	\end{tabular}
\end{table}

The mean of evaluation scores for the predicted masks from the network trained on envelope and B-mode image are $85\%$ and $85\%$ for $DSC$, and $87\%$ and $85\%$ for $F\textsubscript{2}$ score, respectively. The high values in both $DSC$ and $F\textsubscript{2}$ scores indicate that U-Net has a promising structure in segmentation of ultrasound images and is capable in learning the intrinsic features of the simulated data. Furthermore, it shows that the network can learn mappings from the domain of envelope or B-mode image to pixel-level segmentation mask.

\subsection{Predicted Masks for Phantom Data}
\begin{figure}[h!]
 \centering
\begin{minipage}[b]{\linewidth}
 \centering
 \centerline{\includegraphics[width=9cm]{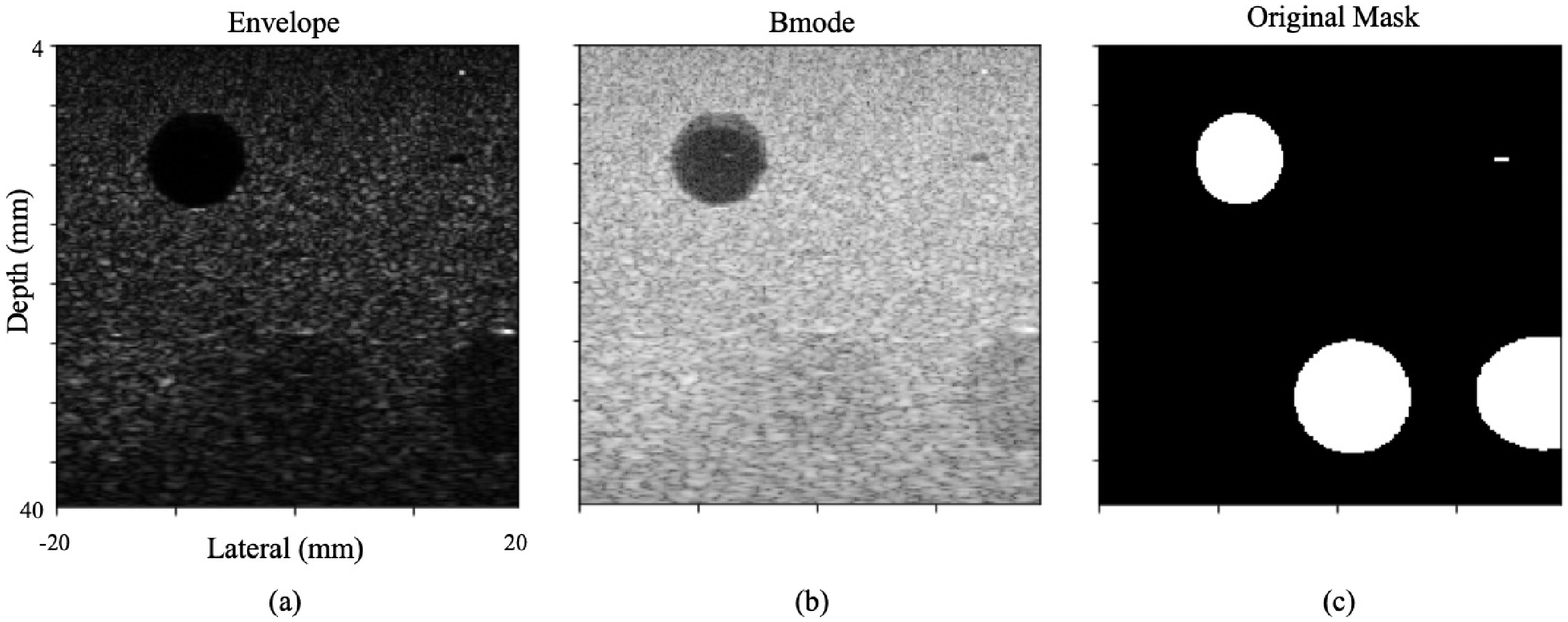}}
\end{minipage}

\begin{minipage}[b]{\linewidth}
 \centering
 \centerline{\includegraphics[width=6.3cm]{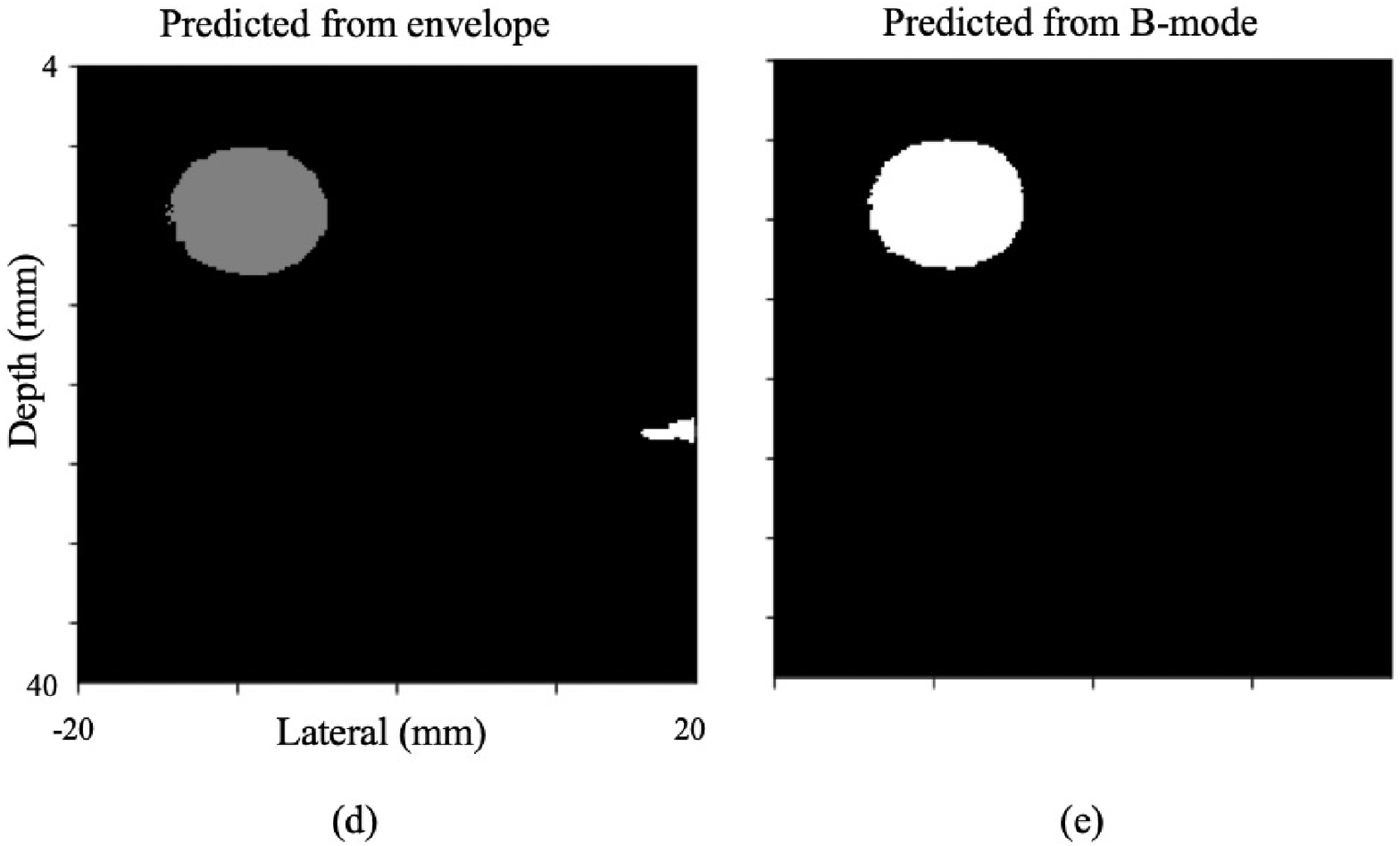}}
\end{minipage}
 \caption{The first row is an example of real tissue-mimicking phantom (a) envelope data, (b) B-mode image, and (c) the ground truth mask. The predicted masks with (d) envelope and (e) B-mode images are shown in the second row.}
 \label{fig:ph}
\end{figure}

An example of the envelope data and B-mode image of our tissue-mimicking phantom is shown in Fig. \ref{fig:ph}. Figures \ref{fig:ph} (d) and (e) show the results of training our network on envelope and B-mode images of simulated data and testing on the phantom data. In this particular example, the phantom data consists of three lesions. In all predicted masks, the anechoic lesion (dark cyst) which is more clearly visible is segmented successfully. The mask derived from envelope data outperforms the B-mode mask as a small portion of hyperechoic lesion is detected. The anechoic lesion predicted from B-mode images is shown in white instead of grey and this is because only one class has been detected.

It is important to highlight that the network has not seen any real images and is fully trained on simulation data. Two conclusions can be made from this observation. First, the Field\_II simulation model is quite similar to real experiments and can be used for training deep learning techniques. Second, the network is not suffering from overfitting and further has learned an efficient representation of US images. 

Table \ref{tab:phscore} presents the quantitative evaluation for phantom data. The mean of evaluation scores for the predicted masks from the network trained on envelope data and B-mode image are $31\%$ and $26\%$ for $DSC$, and $27\%$ and $20\%$ for $F\textsubscript{2}$, respectively.
\begin{table}[h]
	\centering
	\caption{$DSC$ and $F\textsubscript{2}$ scores for real tissue-mimicking phantom data.}
	\label{tab:phscore}
	\small
	\begin{tabular}{ m{4cm}  C{1.5cm} C{1.5cm}}
		\toprule[0.5mm]
		\multicolumn{1}{c}{\bf Predicted Mask} & \multicolumn{1}{c}{$\bf DSC$} & \multicolumn{1}{c}{ $\bf F\textsubscript{\bf2}$}\\
		\toprule[0.5mm]
            	Envelope data & $0.31 \pm 0.23$ & $0.27 \pm 0.2$\\
		\midrule
		B-mode image & $0.26 \pm 0.1$ & $0.2 \pm 0.1$\\
		\midrule
		\midrule
	\end{tabular}
\end{table}

\section{Conclusion}
\label{sec:conclusion}
In this work, we presented the feasibility of training a deep learning architecture on simulated US data and testing on the real tissue-mimicking phantom data. As a consequence, our work offers the use of simulated data as an alternative for datasets with limited training data. Moreover, envelope data outperforms the results derived from training on B-mode images. For future investigations, we aim to test our strategy on \textit{in-vivo} data. We will further examine whether a network trained on simulation data can be fine-tuned using few real experiments.

\section*{Acknowledgement}
We would like to thank NVIDIA for donation of the Titan Xp GPU.

\bibliographystyle{IEEEtran}
\bibliography{refs}

\end{document}